\documentclass[aps,pre,superscriptaddress,nofootinbib,preprintnumbers]{revtex4}

\usepackage{graphicx}
\usepackage{graphics}
\usepackage{amsmath,amssymb}
\usepackage[usenames]{color}
\usepackage{subfigure}
\usepackage{hyperref}
\usepackage{enumitem}
\usepackage{float}

\def\Tr{\,{\rm Tr}\,}

\newcommand{\be}{\begin{equation}}
\newcommand{\ee}{\end{equation}}
\newcommand{\bea}{\begin{eqnarray}}
\newcommand{\eea}{\end{eqnarray}}
\newcommand{\ben}{\begin{enumerate}}
\newcommand{\een}{\end{enumerate}}
\newcommand{\bit}{\begin{itemize}}
\newcommand{\eit}{\end{itemize}}

\newcommand{\la}[1]{\label{#1}}

\newcommand{\Eq}[1]{Eq.~(\ref{#1})}

\newcommand{\Sec}[1]{Sec.~\ref{#1}}

\newcommand{\Fig}[1]{Fig.~\ref{#1}}

\newcommand{\vv}[1]{\mathbf #1}							

\newcommand{\bert}{\raise-0.45mm\hbox{\Large$\Box$}}			

\makeatletter
\newcommand*\bigcdot{\mathpalette\bigcdot@{.5}}
\newcommand*\bigcdot@[2]{\mathbin{\vcenter{\hbox{\scalebox{#2}{$\m@th#1\bullet$}}}}}
\makeatother

\definecolor{BrickRed}{cmyk}{0,0.89,0.94,0.28}					
\definecolor{MidnightBlue}{cmyk}{0.98,0.13,0,0.43}				
\definecolor{DarkGreen}{rgb}{0.100806,0.495968,0.209979}
\definecolor{orange}{rgb}{0.587167,0.354498,0.146197}

\graphicspath {{figures/}}

\begin{document}

\preprint{IFIC/23-29}

\title{The irreversible relaxation of inflation}

\author{Robert Alicki}
\email{robert.alicki@ug.edu.pl}
\affiliation{International Centre for Theory of Quantum Technologies (ICTQT), University of Gda\'nsk, 80-308, Gda\'nsk, Poland}
\author{Gabriela Barenboim}
\email{gabriela.barenboim@uv.es}
\affiliation{Departament de F\'isica Te\`orica and IFIC, Universitat de Val\`encia-CSIC, E-46100, Burjassot, Spain}
\author{Alejandro Jenkins}
\email{alejandro.jenkins@ucr.ac.cr}
\affiliation{International Centre for Theory of Quantum Technologies (ICTQT), University of Gda\'nsk, 80-308, Gda\'nsk, Poland}
\affiliation{Laboratorio de F\'isica Te\'orica y Computacional, Escuela de F\'isica, Universidad de Costa Rica, 11501-2060, San Jos\'e, Costa Rica}
\date{First version: 10 Jul.\ 2023.  This revision: 17 Oct.\ 2024.  Published in Phys.\ Lett.\ B {\bf 866}, 139519 (2025)}

\begin{abstract}
Based on the results of a previous analysis of the Markovian master equation for the irreversible evolution of an open system embedded in de Sitter space \cite{QTdS}, we include in the cosmological Friedmann equations a contribution from the presence of a physical bath at temperature \hbox{$T_{\rm dS} = h / 2 \pi$}, where $h$ is the Hubble parameter.  We show that this provides a mechanism for the irreversible relaxation of the cosmological constant and a graceful exit to inflation, without need for subsequent reheating.  Thermal particle production during inflation gives adiabatic, Gaussian, and approximately scale-invariant cosmological perturbations.  We thus obtain the main features of inflation without any inflaton potential.  To clarify the thermodynamic interpretation of these results, we consider the analogy of this irreversible relaxation to superfluorescence in quantum optics.
\end{abstract}

\maketitle


\section{Introduction}
\la{sec:intro}

The idea, first proposed by Guth in \cite{Guth}, that our Universe is flat and homogenous because it experienced an early de Sitter (dS) phase (i.e., an exponential expansion of space) is now a pillar of modern cosmology.  This ``inflation'' can also account for the primordial perturbations that seeded the subsequent formation of structure in the Universe; see, e.g., \cite{Weinberg} and references thererin.  However, inflation has to a large extent remained what Kolb and Turner called ``a paradigm in search of a model'' \cite{Kolb}.  The most serious obstructions to a satisfactory theory of inflation come from the fine-tuning of the inflaton potential functions and initial conditions needed to explain the observed properties of our Universe.  Moreover, the thermodynamics of dS space remains an obscure and contentious subject in theoretical physics: for recent overviews of some of these controversies; see, e.g., \cite{CLPW, Susskind, deAlwis} and references therein.

In an accompanying article \cite{QTdS}, we studied the dynamics of a localized system weakly coupled to a background of massless quantum fields and embedded in an expanding space.  Such a localized system serves as a quantum thermometer.  The Kubo-Martin-Schwinger (KMS) condition provides a mathematically rigorous and physically meaningful definition of a thermal state, valid both for systems in finite volume with discrete Hamiltonian spectra and for systems in the thermodynamic limit.  According to the KMS condition, the auto-correlation function of any accessible observable $A$ belonging to a certain $C^*$ or von Neumann algebra, must be of the form
\be
\langle A(t) A(s)\rangle = G(t-s) ,
\la{eq:G}
\ee
where the complex-valued function $G(t)$ has the Fourier transform $\tilde G(\omega) \ge 0$ satisfying
\be
\tilde G (-\omega) =  e^{-\omega/k T} \; \tilde{G} (\omega) .
\la{eq:KMS}
\ee

Using the Markovian master equation (MME) for the irreversible evolution of the quantum thermometer, we found in \cite{QTdS} that an observer in the cosmic rest frame with Friedmann-Lema\^itre-Robertson-Walker metric \hbox{$ds^2 = - dt^2 + a^2 (t) \, d \vv x^2$} and constant Hubble parameter \hbox{$h \equiv \dot a / a$} observes a {\it physical} heat bath at the Gibbons-Hawking temperature
\be
T_{\rm dS} = \frac{h}{2 \pi} .
\la{eq:TdS}
\ee
This temperature is only well-defined in the cosmic rest frame: other inertial observers will not see the bath in equilibrium due to the Doppler shifts of the frequencies of the various modes \cite{Sewell}.  We derived the Stefan-Boltzmann law for the contribution of this heat bath to the Universe's energy density,
\be
\rho_{\rm dS} = \sigma h^4 \quad \hbox{for} \quad \sigma \equiv \frac{g_f}{480 \pi^2} ,
\la{eq:SB}
\ee
where $g_f$ is the effective number of massless degrees of freedom.\footnote{Thermal states produced by gravity or by acceleration (e.g., in the Hawking, Unruh, Rindler, or Gibbons-Hawking effects) can be related to pure squeezed states of bosonic fields. Any quantum mixed state can be represented as the partial trace of a pure state of a larger system composed of the original one and an ancilla.  The presence of a causal horizon provides  the decomposition into the observable system and the hidden ancillary one.  This is an instance of the purification procedure used in the rigorous analysis of quantum gases in the thermodynamic limit (the so-called thermofield dynamics).}  Both here and in \cite{QTdS} all quantities are expressed in Planck units, such that \hbox{$G = M_{\rm Pl}^{-2} = \hbar = k_B = c = 1$}.

In \cite{QTdS} we also argued that these results support the claim (made since the 1980s by various theorists on diverse grounds) that quantum effects make dS space unstable in the infrared.\footnote{The earliest such claim appears to have been by Polyakov \cite{Polyakov82}.  Some other references are given in \cite{QTdS}.}  In particular, cosmological particle production at temperature $T_{\rm dS}$ must contribute to the stress-energy tensor $T_{\mu \nu}$ in the Einstein field equations.  This causes the equation of state parameter $w \equiv \rho / p$ (where $\rho = T_{00}$ is the energy density and $p = T_{ii}$ for $i=1,2,3$ is the pressure) to become larger than $-1$, slowing down the Universe's rate of expansion and making the dS background unstable to perturbations \cite{HoHsu}.  In this letter we explore the cosmological consequences of this result.

If \Eq{eq:SB} is extended adiabatically to the case of slowly decreasing $h$, the backreaction of $\rho_{\rm dS}$ on the semiclassical space-time may be obtained {\it nonperturbatively} via the Friedmann equations.  As we will see, this backreaction causes the dS phase to end abruptly as energy is irreversibly transferred from $\rho_{\rm dS}$ into ordinary particles, while $h$ relaxes towards zero.  This offers a ``graceful exit'' to inflation without invoking any inflaton potential function, or even a coherent inflaton field.  Thermal particle production during inflation can also explain the adiabatic, Gaussian, and approximately scale-invariant primordial perturbations that seed the subsequent formation of structure in the Universe.  

A feature of this phenomenological approach is that it allows us to incorporate the effects of cosmological particle production upon the dynamics of the early Universe, without detailed knowledge of the underlying quantum theory of gravity.  To clarify the interpretation of our results we consider the analogy of our model to the phenomenon of superfluorescence in quantum optics, an irreversible process for which the relevant microphysics is well understood.  This analogy might also offer some clues as to how the classical equations of GR emerge from an underlying quantum theory.


\section{Relaxation}
\la{sec:relax}

In light of \Eq{eq:SB}, we write the total energy density of the Universe as $\rho_{\rm dS}$ plus another component $\rho_r$ corresponding to regular particles (radiation and matter),
\be
\rho = \rho_{\rm dS} + \rho_r = \sigma h^4 + \rho_r ,
\la{eq:rhos}
\ee
where $\rho_{\rm dS}$ does not dilute as the Universe expands with constant $h$.  The corresponding pressure $p_{\rm dS}$ must therefore obey the equation of state $p_{\rm dS} / \rho_{\rm dS} = -1$.\footnote{The equation of state of any system in equilibrium depends on its boundary conditions.  For a gas in a fixed but controllable volume $V$, the equation of state is obtained from the formula $p = -\partial F / \partial V$, where $p$ is the pressure and $F$ the free energy. In this case we have taken $F = \rho_{\rm dS} V$, with the boundary condition imposed by fixing the vacuum's energy density $\rho_{\rm dS}$.}  Thus, the cosmological pressure $p$ is given by
\be
p = p_{\rm dS} + p_r = - \sigma h^4 + w_r \rho_r ,
\la{eq:ps}
\ee
where $w_r = p_r / \rho_r$ is the equation of state for the regular particle content of the Universe.  Note that, because the $\rho_{\rm dS}$ term scales as $h^4$ (rather than as $h^2$), it is {\it not} a cosmological constant term.  Rather, it corresponds to a gas formed and maintained by cosmological particle production, which is a gravitational process \cite{HoHsu}.  As we will see in detail, this allows $\rho_{\rm dS}$ to relax towards zero while $\rho_r$, which starts at zero, grows until it overtakes $\rho_{\rm dS}$, thus reheating the Universe.

By combining Eqs.\ \eqref{eq:rhos} and \eqref{eq:ps} with the Friedmann equations, we take into account the back reaction of cosmological particle production on the semiclassical space-time in a non-perturbative way and without requiring detailed knowledge of the correct quantum theory of gravity.  For this, it is only necessary to assume that the thermalization to the instantaneous temperature $T_{\rm dS} \propto h$ (which is a gravitational effect, associated with cosmological particle production) is fast enough that the Stefan-Boltzmann law of \Eq{eq:SB} can be applied adiabatically at each instant, even if $h$ is not constant.  It is easy to check, from the solution to $h(t)$ that we will obtain from our model, that $|\dot h / h|$ remains well below 1 in Planck units, implying that the adiabatic approximation is consistent.

\begin{figure}[t]
	\includegraphics[height=0.25 \textwidth]{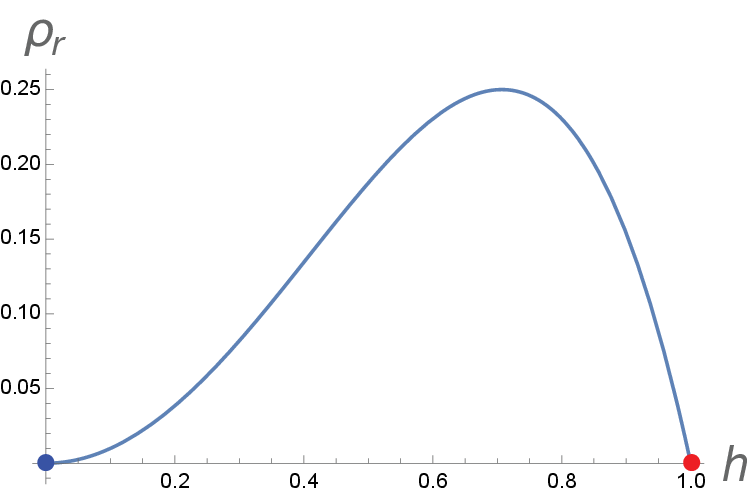}
\caption{Energy density $\rho_r$ in the regular particle content of the Universe (radiation and matter) as a function of the Hubble parameter $h$, according to \Eq{eq:1FE}.  The horizontal scale is in units of $h_{\rm BD} = \sqrt{3 / 8 \pi \sigma}$ [\Eq{eq:hBD}].  The vertical scale is in units of $\sigma h_{\rm BD}^4$.  The two solutions with $\rho_r = 0$ are: $h=0$ (Minkowski vacuum, marked in blue) and $h=h_{\rm BD}$ (Bunch-Davies vacuum, marked in red).\la{fig:rho-h}}
\end{figure}

Combining the first Friedmann equation ($h^2 = 8 \pi \rho / 3$) with \Eq{eq:rhos}, we find that
\be
\rho_r = \frac{3}{8\pi }h^2 - \sigma h^4 .
\la{eq:1FE}
\ee
Equation \eqref{eq:1FE} has two solutions with $\rho_r = 0$ (see \Fig{fig:rho-h}).  One is the static and empty Universe ($h = 0$), while the other, with
\be
h = h_{\rm BD} \equiv \sqrt{\frac{3}{8 \pi \sigma}} = 6 \sqrt{\frac{5 \pi}{g_f}} ,
\la{eq:hBD}
\ee
which we identify with the Bunch-Davies vacuum, as it corresponds to a state of dS space with no particle content.  Note that, for $g_f \sim 10^2$ (as in the Standard Model) the value of $h_{\rm BD}$ is at the Planck scale (i.e., $h_{\rm BD} \sim 1$ in the natural units that we are using).  As we shall see, this implies that the dS phase ends naturally by reheating at a Planck-scale temperature.

Taking the second Friedmann equation ($\dot h = - 3 h^2/2 - 4\pi p$) with $p$ given by \Eq{eq:ps} and $\rho_r$ by \Eq{eq:1FE}, we find that
\be
\dot h = - 4 \pi\sigma (1+w_r) h^2 \left( \frac{3}{8\pi\sigma}- h^2 \right) = - \frac 3 2 (1 + w_r) \left( \frac{h}{h_{\rm BD}} \right)^2 \left( h_{\rm BD}^2 - h^2 \right) .
\la{eq:2Fried1}
\ee
Note that $h^2 > h_{\rm BD}^2$ is forbidden, as it would imply $\rho_r < 0$ in \Eq{eq:1FE}.  Equation \eqref{eq:2Fried1} has an unstable fixed point at $h = h_{\rm BD}$ (the Bunch-Davies vacuum) and a stable fixed point at $h = 0$ (the Minkowski vacuum).  If the Universe is initially in the Bunch-Davies vacuum, a small perturbation can trigger the relaxation $h \to 0$, accompanied by particle production ($\rho_r > 0$) and a graceful exit from inflation, without the need for a separate phase of reheating after inflation.  It should be stressed that, whereas the consistency of slow-roll dynamics imposes a condition $h \lesssim 10^{-5} M_{\rm Pl}$ in such models of inflation (see, e.g., ch.\ 10 in \cite{Weinberg}), in our case $h(0)$ can be at the natural scale \hbox{$M_{\rm Pl} \sim 10^{19}$ GeV}.

Let us define the characteristic time $t_c$ such that
\be
\rho_c \equiv \rho_{\rm dS} (t_c) = \rho_r (t_c) .
\la{eq:rho-eq}
\ee
This is the only time in the history of the Universe when relativistic matter and radiation are in equilibrium with the temperature $T_{\rm dS}$.  Moreover, $\rho_r (t_c)$ and $\dot h^2 (t_c)$ are the maximum values of these quantities as functions of time, as shown in \Fig{fig:history}.  Using \Eq{eq:1FE} one may compute
\be
h(t_c) = \left(\frac{3}{16 \pi \sigma} \right)^{1/2} = \frac{h_{\rm BD}}{\sqrt 2} ,
\la{eq:maxh}
\ee
which implies that
\be
\rho_c \equiv \rho_r(t_c) = \rho_{\rm dS}(t_c) = \left(\frac{3}{16\pi}\right)^2 \frac{1}{\sigma} .
\la{eq:maxdensity}
\ee
The equilibrium temperature is 
\be
T_c \equiv T_{\rm dS} (t_c) = \frac{h(t_c)}{2 \pi} = \frac{h_{\rm BD}}{2 \pi \sqrt 2} .
\la{eq:Tc}
\ee
The standard hot Big-Bang cosmology starts at time $t_c$, when the Universe enters into a radiation-dominated state.  Thus, the temperature $T_c$ of \Eq{eq:Tc} plays the role of the reheating temperature in standard inflationary cosmology, except that it now takes a much higher (Planck scale) value.  The precise value of $T_c$ depends on the value of $g_f$ in \Eq{eq:hBD}.

\begin{figure}[t]
	\includegraphics[height=0.4 \textwidth]{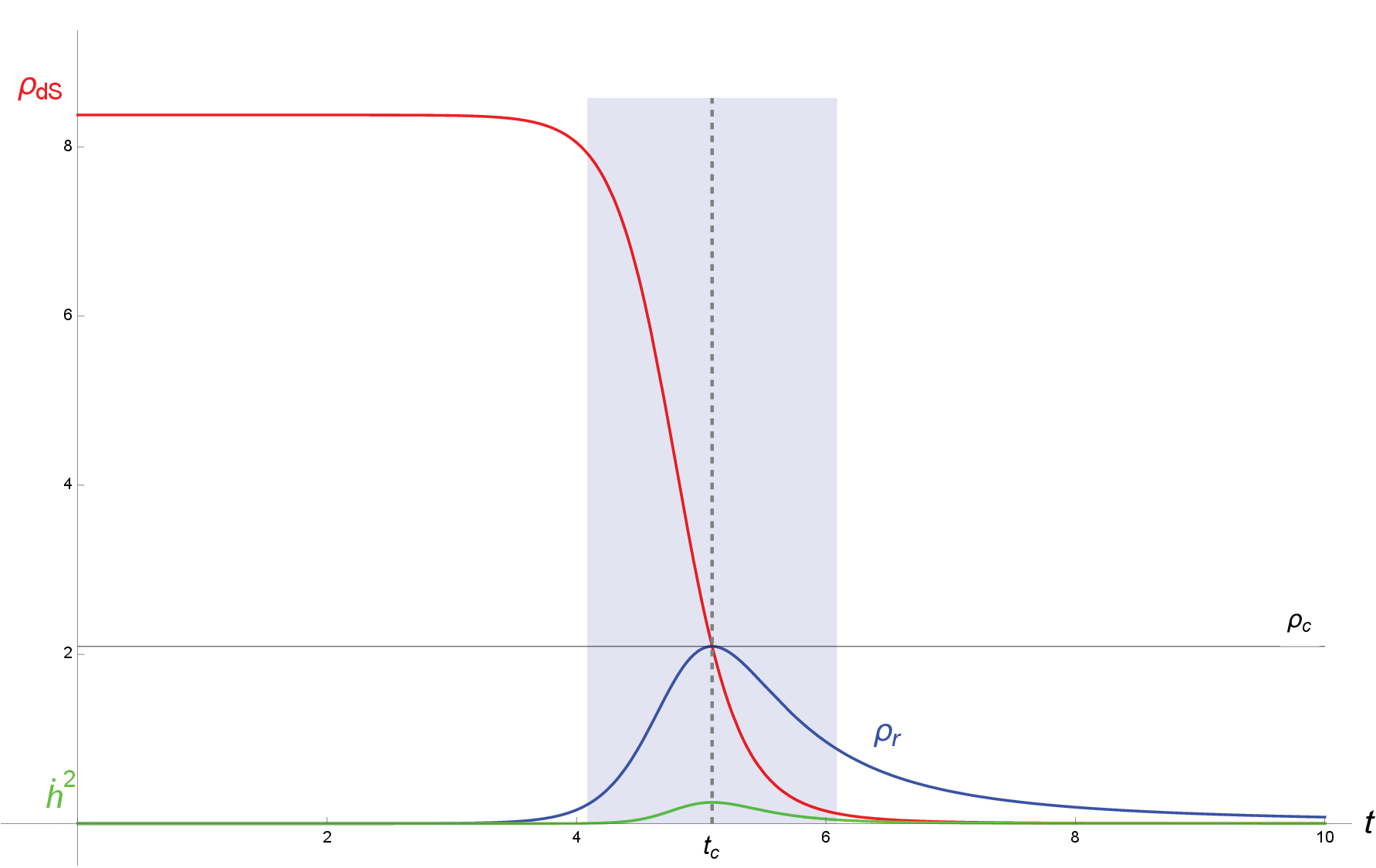}
\caption{Cosmological history based on the solution to \Eq{eq:2Fried1} for initial condition $h(0) = h_{\rm BD}(1 - 10^{-9})$.  The quantities plotted are $\rho_{\rm dS}(t)$ [\Eq{eq:SB}, in red] and $\rho_r(t)$ [\Eq{eq:1FE}, in blue], and $\dot h^2(t)$ (in green).  The horizontal scale is in units of Planck time.  The vertical scale is in units of $\sigma h_{\rm BD}^4$ for the densities $\rho_{\rm dS}$ and $\rho_r$, and $h_{\rm BD}^4$ for $\dot h^2$.  The characteristic time $t_c$, defined in \Eq{eq:rho-eq}, is marked by the dashed vertical line, while the light blue shading indicates the interval \hbox{$| t - t_c | < 1$}, during which particle production is important.  The value of $\rho_c$ in \Eq{eq:rho-eq} is marked by the solid horizontal line.\la{fig:history}}
\end{figure}

\section{Analogy to superfluorescence}
\la{sec:SF}

To help clarify the thermodynamic interpretation of \Eq{eq:2Fried1}, we will consider here the analogy to {\it superfluorescence}.  This is a process in quantum optics in which a large number of initially excited atoms relax spontaneously to their ground states by producing a strong and narrow pulse of monochromatic radiation.  Such a phenomenon was predicted by Dicke in 1954 \cite{Dicke} and has since attracted considerable interest from both theorists and experimentalists in quantum optics.\footnote{Dicke introduced the term ``super-radiant'' in \cite{Dicke}.  Bonifacio and Lugiato later made a distinction between superradiance, in which the initial state is highly coherent, and superfluorescence, in which it is not \cite{SF}.  In that sense, the relevant analogy of our cosmological model is to superfluorescence.  Gross and Haroche use ``superfluorescence'' as a synonym of ``superradiance'' in \cite{Gross-Haroche}, while Alicki and Lendi use only the latter term in \cite{Alicki-Lendi}.  However, some later authors have distinguished between the two; see, e.g., \cite{SF-perovskite}.  Here we prefer {\it superfluorescence} not only because the initial state in our cosmological model is incoherent, but also because in high-energy physics {\it superradiance} is commonly applied to a different process, in which the kinetic energy of a moving heat bath (e.g., a Kerr black-hole) is irreversibly transformed into non-thermal radiation; see, e.g., \cite{rotatingbath, BCP} and references therein.}  The analogy between the irreversible relaxation of inflation and superfluorescence will also help us to interpret the choice of initial condition in the cosmological \Eq{eq:2Fried1}.

The simplified treatment of superfluorescence given here is based on \cite{Alicki-Lendi}.  Superfluorescence can be idealized in terms of $N$ identical two-level ``atoms'', each with atomic angular frequency $\omega_A$, which do not interact with each other, but which are collectively coupled to an electromagnetic field at zero temperature.  In the simplest theoretical description, valid in the large $N$ limit, we assume a product structure for the initial density matrix of the atomic ensemble, which corresponds to Boltzmann's {\it Stosszahlansatz}.  Using the standard methods to derive the MME for an open quantum system, we may show that the product structure of the atomic state $\hat \rho_N (t)$ is preserved in time:
\be
\hat \rho_N (t) = \hat \rho (t) \otimes \hat \rho (t) \otimes \cdots \otimes \hat \rho (t) \quad \hbox{for} \quad
\hat \rho (t) = \begin{pmatrix} p(t) & z(t) \\ z^\ast(t) & 1-p(t) \\ \end{pmatrix} , 
\la{eq:productstate}
\ee 
with consistency conditions
\be
0 \leq p(t) \leq 1, \quad  |z(t)|^2 \leq p(t) [1 - p(t) ] .
\la{eq:p-z}
\ee
Here, $p(t)$ is the fraction of the atoms that are in their excited state at time $t$.  This is proportional to the average energy per atom,
\be
\epsilon_A (t) = \hbar \omega_A p(t) .
\la{eq:epsA}
\ee
Meanwhile, the complex quantity $z(t)$ is the atomic ``quantum coherence''.  The evolution of the system is governed by the nonlinear equations of motion
\begin{align}
\dot p &= - \gamma_N |z|^2 , \la{eq:MDM-p} \\
\dot z &= - i \omega_A  z +  \gamma_N \left(p - \frac{1}{2}\right) z , \la{eq:MDM-z}
\end{align}
where $\gamma_N = N \gamma_e$ is the rescaled, macroscopic decay rate given by the single atom spontaneous emission rate $\gamma_e$.

Even though the evolution described by Eqs.\ \eqref{eq:MDM-p} and \eqref{eq:MDM-z} is thermodynamically irreversible (which results from tracing over the Hilbert subspace of the electromagnetic field) it can be shown that, for the reduced description in terms of the single-atom density matrix $\rho$, a pure initial state state [$\rho^2 (0) = \rho (0)$] remains pure, so that
\be
\Tr{\rho^2 (t)} = 1 ,
\la{eq:trace}
\ee
for all $t$.  Equation \eqref{eq:trace} implies that
\be
|z(t)|^2  =  p(t) - p^2 (t) .
\la{eq:pure}
\ee 
Combining this with \Eq{eq:MDM-p} gives us a simple kinetic equation for $p(t)$:
\be
\dot p = -\gamma_N p (1 - p) .
\la{eq:dp}
\ee
Equation \eqref{eq:dp} is analogous to the cosmological \Eq{eq:2Fried2} in that it has two fixed points: a stable one at $p=0$ and an unstable one at $p=1$.  Note also that \Eq{eq:pure} has a similar structure to the cosmological energy density of regular particles of \Eq{eq:1FE}.  The evolutions of $p^2(t)$ and $|z(t)|^2$, as shown in \Fig{fig:superfluorescence}, therefore resemble the behaviors of $\rho_{\rm dS} (t)$ and $\rho_r(t)$, respectively.

\begin{figure}[t]
	\includegraphics[height=0.3 \textwidth]{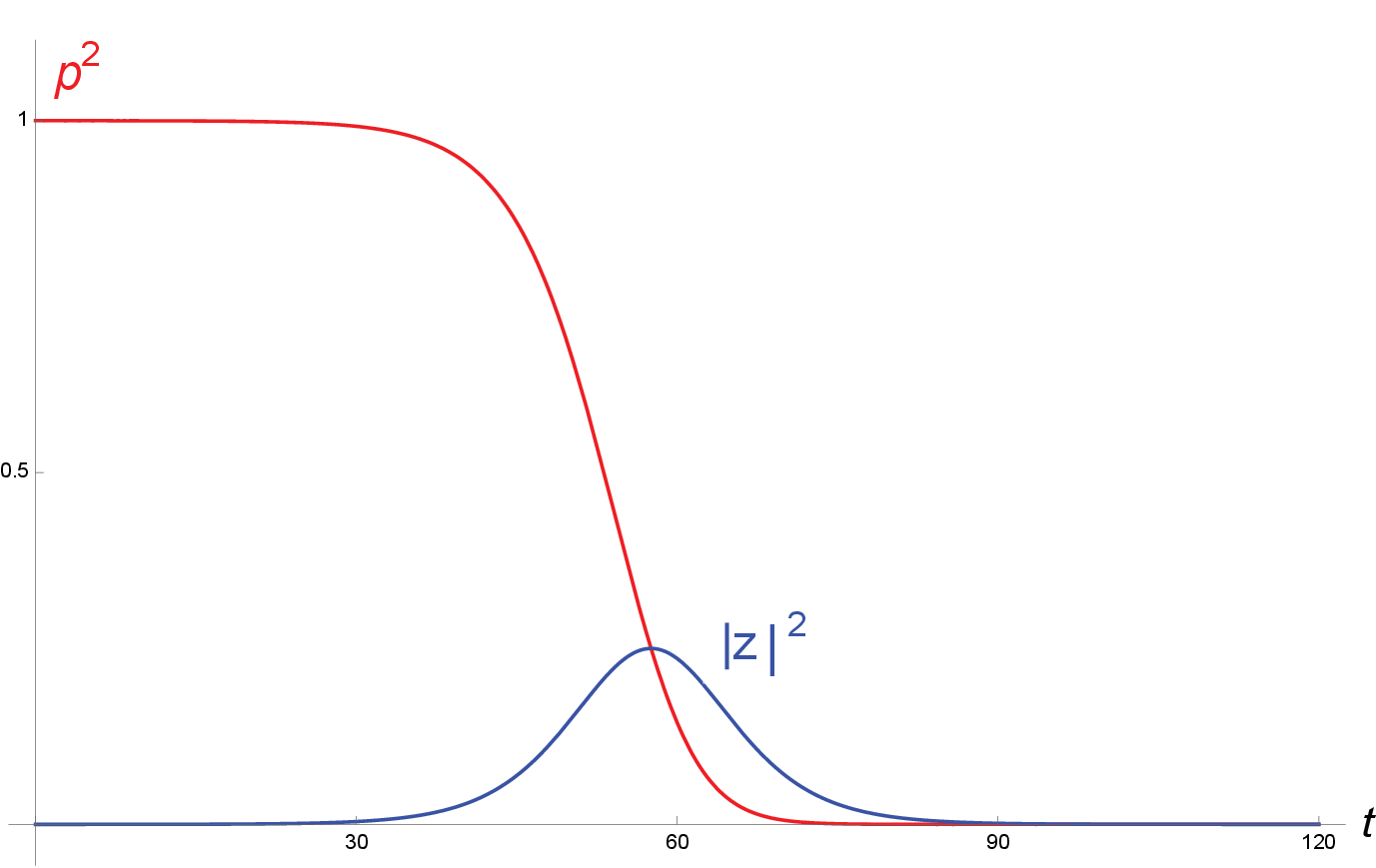}
\caption{Dynamics of superfluorescence, illustrated by the solution for $p^2(t)$ (in red) and the corresponding $|z(t)|^2$ (in blue), according to Eqs.\ \eqref{eq:pure} and \eqref{eq:dp}, with decay rate $\gamma_N = 0.2$ and initial condition $p(0) = 1 - 10^{-5}$.\la{fig:superfluorescence}}
\end{figure}

The analogy between superfluorescence and the irreversible relaxation of inflation proposed here may be deeper than a mere mathematical similarity of the relevant differential equations.  Equations \eqref{eq:pure} and \eqref{eq:dp} describe a process of collective spontaneous emission, in which an initial state with almost all of the atoms excited relaxes irreversibly towards the ground state.  Those equations can be obtained from the unitary evolution of the total physical system (atoms plus electromagnetic field) using the standard MME along with a semiclassical approximation \cite{Alicki-Lendi}.  The result is a nonlinear but purity-preserving evolution of the reduced density matrix for a single atom, expressed by the kinetic \Eq{eq:dp}.  This is at the same level of description as the Boltzmann equation for a gas of particles, except that it also involves macroscopic quantum coherence effects, captured by $z(t)$.  The model of the irreversible relaxation of inflation based on \Eq{eq:2Fried2} could have a similar thermodynamic interpretation, so that one might hope to derive it in the future from a more fundamental theory of quantum gravity in terms of a large and highly symmetric collection of qubits, in the spirit of ``it from qubit''.

In such an interpretation, the Friedmann equations would not be classical field equations with classical sources, but rather a mean-field approximation (akin to the time-dependent Hartree or Hartree-Fock equations) for the irreversible dynamics of an open quantum system with some yet unknown fundamental degrees of freedom.  Those mean-field equations do not include all quantum multiparticle correlations, but they reproduce the leading-order quantum features in the regime of large number of particles.

Note that, for superfluorescence in the mean-field limit ($ N\to\infty$), an initially pure product state remains a pure product, so that the entropy per atom vanishes during the evolution.  However, finite-size effects (such as those considered in \cite{mean-field}) produce a residual entropy that scales more slowly than $O(N^1)$.  That property may be analogous to the holographic principle in quantum gravity \cite{holographic}. Note that such approximate purity preservation does not contradict the usual picture of the emergence of classicality through decoherence, since in quantum physics a pure state of the total system is compatible with high-entropy states for local subsystems.


\section{Initial condition}
\la{sec:initial}

Changing variables $t \to h_{\rm BD} t$ and $h \to h / h_{\rm BD}$, \Eq{eq:2Fried1} becomes
\be
\dot h = - \frac 3 2 (1+ w_r) \left( h^2 - h^4 \right) .
\la{eq:2Fried2}
\ee
We take $h(0)$ in \Eq{eq:2Fried2} very close to $1$, so that the number of e-folds is equal to the time elapsed from $t = 0$ to the end of inflation.  Substituting $h(t) = 1 - \epsilon(t)$ into \Eq{eq:2Fried2}, we have
\be
\dot \epsilon = \frac 3 2 (1 + w_r) \left[ \left( 1 - \epsilon \right)^2 - \left(1 - \epsilon \right)^4 \right] = 3 (1 + w_r) \epsilon + O \left( \epsilon^2 \right) .
\la{eq:eps}
\ee
If $w_r$ is a constant during inflation, the solution to \Eq{eq:eps} will behave initially as \hbox{$\epsilon(t) \simeq \epsilon(0) e^{3(1+w_r)t}$}.  Inflation ends when the condition $\epsilon (t) \ll 1$ no longer holds.  Therefore, to get 60 e-folds we need
\be
\epsilon (0) \ll e^{-180 (1+w_r)} .
\la{eq:60ef}
\ee

Since $w_r$ gives the equation of state for the contents of the Universe (other than the bath), taking $w_r = -1$ in \Eq{eq:60ef} would make inflation last forever.  However, if this were a Bose-Einstein condensate (BEC), heating by the background radiation at $T_{\rm dS}$ could cause a phase transition, with the equation of state evolving from $w_r = -1$ to $w_r = 1/3$.  The number of e-folds would depend on the magnitude of the coupling between the initial BEC and the dS bath.  The energy density of the BEC could be taken constant, without tuning any potential.

Alternatively, we could take $w_r = 1/3$, corresponding to no coherent inflaton at all.  Equation \eqref{eq:60ef} then becomes \hbox{$\epsilon(0) \ll e^{-240}\simeq 10^{-104}$}.  This may seem an unacceptable fine tuning of the initial condition, but a more compelling interpretation is suggested by the analogy to superfluorescence that we have considered above.

In superfluorescence, the initial state of the system of $N$ excited atoms is stationary in the semiclassical limit $N \to \infty$ and an initial fluctuation is needed to trigger superfluorescent emission.  Let us suppose that the classical intensive variable  $\epsilon = 1 - h \geq 0$ in \Eq{eq:eps} were the mean value of the square of a collective quantum observable,
\be
\epsilon = \langle \hat e_N ^2 \rangle , \quad \hbox{for} \quad  \hat e_N = \frac 1 N \sum_{j=1}^N \hat{\xi}_j .
\la{eq:epsN}
\ee
One may think of this in terms of the large ensemble of $N$ qubits underlying the semi-classical space-time. We compute $\epsilon (0)$ for the state corresponding to the maximum classical value of the Hubble parameter [$h(0) = 1$ in \Eq{eq:2Fried2}].  If the $N$-atom state is approximately a product state or a coherent state, we have that
\be
\epsilon(0) \simeq \frac{1}{N^2} \sum_{j,k =1}^N \langle \xi_j \xi_k \rangle = \frac{1}{N^2} \sum_{j=1}^N \langle \xi^2_j  \rangle = O \left( N^{-1} \right) ,
\la{eq:hgrav}
\ee
assuming that $\langle \xi_k\rangle = 0$ and $\langle \xi^2_k\rangle = O(1)$.  Combining Eqs.\ \eqref{eq:60ef} and \eqref{eq:hgrav}, the condition for enough e-folds translates to $N \gg 10^{104}$.  One may compare this to estimates of various large numbers characterizing the visible Universe: e.g., $10^{90}$ for the number of photons, and $10^{80}, 10^{104}, 10^{122}$ or $10^{124}$ for the total entropy or ``number of bits'' (see \cite{entropy} and references therein).


\section{Perturbations}
\la{sec:perturbations}

Rather than interpreting primordial cosmological perturbations as resulting from quantum vacuum fluctuations (which in statistical physics are renormalized to zero by normal ordering of the operators in the fields), we could regard them as thermal fluctuations.  This is akin to ``warm inflation'', except that in that literature the thermal fluctuations are obtained from classical Langevin equations (see \cite{warm} and references therein), while in our case the fluctuations are given by quantum noise spectra that can be computed from the MME for both bosons and fermions.  Moreover, the temperature of these spectra is given directly by $h$ and is therefore independent of the particle content or interactions.  The bosonic spectrum ${\cal P}_{\rm th} (\omega)$ derived in \cite{QTdS} is
\be
\mathcal{P}_{\rm th}(\omega)= \frac{4 \pi \omega}{(e^{\beta\omega}-1)} \simeq 2 h .
\la{eq:Pth}
\ee
for $\beta = 2 \pi / h$, where the approximation holds for $\omega \ll h$.  It is clear that such perturbations will be adiabatic, Gaussian, and approximately scale invariant.

Since we do not invoke any slow-roll potential $V(\phi)$ or even a coherent inflaton $\phi$, detailed analysis of the cosmological observables resulting from the irreversible relaxation described by \Eq{eq:2Fried1} requires a new calculational technology that we must leave for future research.  At this stage, we can say that our model probably implies an unobservable tensor-to-scalar ratio $r$, since the power of the perturbations is distributed equally in all of the $g_f$ massless field degrees of freedom, including gravitons.

If \Eq{eq:Pth} is re-expressed in terms of comoving wavenumber $k$ as a spectrum $P_k \propto h$ and evaluated at the time of horizon crossing for each mode [$k = a h$, so that $\dot k = a (h^2 + \dot h)$], we get that the tilt of the cosmological perturbations
\be
n_s - 1 = \left. \frac{d \ln P_k}{d \ln k} \right|_{k = a h} = \left. \frac{k \dot P_k}{\dot k P_k} \right|_{k = a h} = \frac{\dot h}{h^2 + \dot h}
\la{eq:tilt}
\ee
is small and negative (``red'').  We expect that fixing the tilt will fix $T_c$ in \Eq{eq:Tc}, and therefore $g_f$ in \Eq{eq:SB}.  The value of $n_s - 1$ obtained from the current Planck satellite data seems to be consistent with $g_f \simeq 600$.


\section{Discussion}
\la{sec:discussion}

Previous authors have argued that dS space is dynamically unstable and that this may explain how inflation ends and why the cosmological constant is currently so small compared to the natural scale $M_{\rm Pl}^4$ (for some of the more recent arguments see, e.g., \cite{Polyakov12, Dvali, HoHsu, Brandenberger}).  However, the thermodynamics of dS space has remained poorly understood.  In \cite{QTdS} we showed, using the analytical methods of quantum thermodynamics, that $T_{\rm dS} = h / 2 \pi$ is the temperature of a {\it physical} bath.  In particular, this temperature is only well defined in a preferred frame of reference (the ``cosmic rest frame''), in which a localized observer sees the dS vacuum as a state in thermal equilibrium.  This is the same frame of reference that, later in the history of the Universe, will become the rest frame of the cosmic microwave background.  The presence of such a preferred frame is a general feature of temperature in quantum physics \cite{Sewell}.  Our results support the conclusion that the classical isometries of dS space are anomalous \cite{trouble}.

In most of the literature on inflationary cosmology, any thermal effects associated with $T_{\rm dS}$ are taken to be negligible relative to the slow-roll dynamics of the inflation field $\phi$, governed by the form of its potential $V(\phi)$.  This is so by design: consistency of the slow-roll dynamics requires $T_{\rm dS}$ to be well below $M_{\rm Pl}$ \cite{Weinberg}.  But the model presented here has no inflaton at all.  Instead, the inflationary dynamics are {\it wholly} driven by thermal effects associated with $T_{\rm dS}$, which starts at the Planck scale and decays irreversibly towards zero while reheating the Universe.  By including the contribution from the Stefan-Boltzmann law for the energy density associated with $T_{\rm dS}$ in the Friedman equations, we arrived at a phenomenologically consistent picture of the irreversible relaxation of inflation.

Such an irreversible relaxation process is somewhat analogous to that of a superheated liquid (the dS bath) after an initial perturbation that triggers spontaneous boiling (analogous to cosmological particle production) while reducing the temperature of the evaporating liquid (analogous to $\dot h < 0$).   A more precise analogy can be drawn to superfluorescence in quantum optics, as we considered in detail in \Sec{sec:SF}.  We regard this analogy as useful in part because it seems to point towards how the classical field equations of GR emerge from an underlying quantum theory (as yet unknown) in the spirit of ``it from qubit''.

In order to obtain enough e-folds, our model requires an initial value $h(0)$ that is very close to the unstable fixed point of \Eq{eq:2Fried1} at $h = h_{\rm BD}$.  Determining whether or not this is a natural initial condition would require a better understanding of the underlying quantum theory, but it is encouraging that $h = h_{\rm BD}$ is one of only two solutions of \Eq{eq:1FE} for an empty Universe ($\rho_r = 0$), the other corresponding to the Minkowski vacuum ($h =0$).  The analogy to superfluorescence allowed us to argue in \Sec{sec:initial} that the smallness of $h_{\rm BD} - h(0)$ might be naturally explained as a quantum fluctuation in a universe with an entropy as large as the one that our Universe is believed to have.

Based on these results we are hopeful that, by treating the expansion of the Universe as an explicitly irreversible process using the modern analytical methods of quantum thermodynamics, one may arrive at a considerable simplification of the dynamics of the early Universe.  Further research is needed to establish in detail what this qualitatively new theoretical implementation of inflation predicts for observable quantities.  We consider this investigation as worth pursuing, since it holds the promise of realizing Lema\^itre's original vision, according to which the pure quantum state of the Universe is unstable and therefore decays irreversibly into the mixed, non-equilibrium state that we observe \cite{Lemaitre}.  \\


{\bf Acknowledgements}: We thank Steve Hsu, Jonathan Oppenheim, Enrico Pajer, and Fernando Quevedo for discussions.  RA and AJ were supported by the International Research Agendas Programme (IRAP) of the Foundation for Polish Science (FNP), with structural funds from the European Union (EU).  GB was supported by the Spanish Grants No.\ PID2020-113334GB-I00/AEI/1013039/501100011033 and No.\ CIPROM/2021/054 (Generalitat Valenciana). GB has also received support from the EU's Horizon 2020 research and innovation program under the Marie Sk{\l}odowska-Curie Grant Agreement No.\ 860881-HIDDeN.


\end{document}